\documentclass{emulateapj}

\shortauthors{Burgh et al.}
\shorttitle{Recombination Ghosts in VPH Gratings}

\begin{document}

\title{Recombination Ghosts in Littrow Configuration: Implications for
  Spectrographs Using Volume Phase Holographic Gratings}

\author{Eric B. Burgh\altaffilmark{1,2}, 
  Matthew A. Bershady\altaffilmark{3}, 
  Kyle B. Westfall\altaffilmark{3}, 
  Kenneth H. Nordsieck\altaffilmark{1}} 
\email{ebb@sal.wisc.edu}

\altaffiltext{1}{Space Astronomy Laboratory, University of
  Wisconsin-Madison, 1150 University Avenue, Madison, WI 53706}

\altaffiltext{2}{Current address: Center for Astrophysics and Space
  Astronomy, University of Colorado, 389-UCB, Boulder, CO 80309}

\altaffiltext{3}{Astronomy Department, University of
  Wisconsin-Madison, 475 North Charter Street, Madison, WI 53706}

\received{27 June 2007}
\revised{2 August 2007}
\accepted{2 August 2007}

\begin{abstract} 
We report the discovery of optical ghosts generated when using Volume
Phase Holographic (VPH) gratings in spectrographs employing the
Littrow configuration. The ghost is caused by light reflected off the
detector surface, recollimated by the camera, recombined by, and
reflected from, the grating and reimaged by the camera onto the
detector. This recombination can occur in two different ways.  We
observe this ghost in two spectrographs being developed by the
University of Wisconsin - Madison: the Robert Stobie Spectrograph for
the Southern African Large Telescope and the Bench Spectrograph for
the WIYN 3.5~m telescope. The typical ratio of the brightness of the
ghost relative to the integrated flux of the spectrum is of order
$10^{-4}$, implying a recombination efficiency of the VPH gratings of
order $10^{-3}$ or higher, consistent with the output of rigorous
coupled wave analysis. Any spectrograph employing VPH gratings,
including grisms, in Littrow configuration will suffer from this
ghost, though the general effect is not intrinsic to VPH gratings
themselves and has been observed in systems with conventional gratings
in non-Littrow configurations. We explain the geometric configurations
that can result in the ghost as well as a more general prescription
for predicting its position and brightness on the detector. We make
recommendations for mitigating the ghost effects for spectrographs and
gratings currently built. We further suggest design modifications for
future VPH gratings to eliminate the problem entirely, including
tilted fringes and/or prismatic substrates. We discuss the resultant
implications on the spectrograph performance metrics.
\end{abstract}


\section{Introduction}

Modern astronomical spectrographs are being designed and built to
maximize efficiency in all possible ways.  CCD quantum efficiencies
are nearing 100\% and coatings, both reflection and anti-reflection,
are close to their performance limits as well. The introduction to
astronomy of volume phase holographic (VPH) gratings
\citep{Barden00,Baldry04} has further increased routine efficiency by
as much as factors of two.

A VPH grating consists of a thin (3-30~$\mu$m) layer of dichromated
gelatin (DCG) sandwiched between glass substrates.  Through exposure
to a laser interferogram, the DCG's refractive index is modulated in a
sinusoidal pattern, yielding ``fringes'', functionally analogous to
grooves in a ruled grating, with the principle distinction that the
fringes are in a volume not on a surface. With appropriate orientation
of the fringe plane, the VPH grating can function either in
transmission or reflection.  The advantage of VPH gratings relative to
conventional surface-relief gratings is their high efficiency (up to
90\%), large super-blaze (i.e., good efficiency over a broad range of
tunable central wavelengths), low scattered light, and transmissivity
as well as reflectivity.  Transmissivity permits more compact
spectrograph designs, particularly for large incidence-angle (i.e.,
high dispersion) setups, which allows for more optimum pupil
placement, and hence less vignetting.  VPH gratings are becoming
quite common and are being designed for, or are already being used by, a
large number of spectrographs, including some that have been
retrofitted with VPH grisms.

The Department of Astronomy and the Space Astronomy Laboratory at the
University of Wisconsin - Madison are developing two spectrographs
that will also take advantage of this new technology: The Robert
Stobie Spectrograph (RSS), formerly called the Prime Focus Imaging
Spectrograph, for the Southern African Large Telescope (SALT) and an
upgrade for the Bench Spectrograph on the WIYN\footnote{The WIYN
Observatory is a joint facility of the University of
Wisconsin~-~Madison, Indiana University, Yale University, and the
National Optical Astronomy Observatories.} 3.5~meter telescope.

\begin{figure*}
\epsscale{1.2}
\plotone{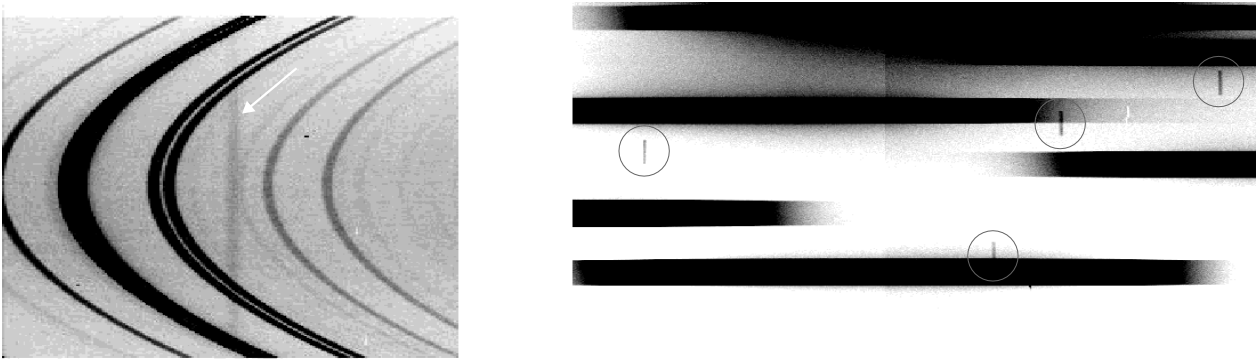}
\caption{Left: SALT/RSS detection of Littrow ghost (straight vertical
line in center) observed in a longslit arc-lamp spectrum with a 2300
lines~mm$^{-1}$ grating at $\alpha=50\degr$.  This is the highest
grating angle for RSS and as such the line curvature is maximal,
although in this image the vertical axis has been compressed by a
factor of eight to further accentuate the curvature of the dispersed
spectral lines, making the Littrow ghost more obvious.  Right:
Close-up view of Littrow ghosts (circle) from a continuum lamp taken
though a multi-object slitmask on RSS with a 3000 lines~mm$^{-1}$
grating at $\alpha\sim47\degr$. Each ghost looks like an image of
a slitlet on the multi-object slitmask and is situated opposite the
center of field from the Littrow wavelength in its corresponding
spectrum (not in this figure because of the close-up).  Both examples
use VPH gratings in first order. Wavelength increases from left to
right.
\label{fig1}}
\end{figure*}

While testing the RSS VPH gratings during the assembly and integration
phase, we identified a spurious feature that appeared near the center
of the CCD array for all gratings and grating angles (see Figure
\ref{fig1}).  We noted that it moved by an angle exactly twice that of
any grating rotations off the nominal, Littrow configuration and thus
determined that it must arise from a reflection off of, or internal
to, the grating.  Subsequent measurements during the commissioning of
VPH gratings for the Bench Spectrograph (see Figure \ref{fig2}) showed
the ghost to not be a feature limited to RSS or one VPH grating
manufacturer.

We eventually hypothesized that the ghost was caused by light
reflected off of the detector plane, recollimated by the camera,
recombined by the VPH grating, and reimaged onto the detector. We
called it the ``Littrow ghost'' because it is a natural consequence of
using a grating at Littrow configuration and is not unique to VPH
gratings. Although the presence and nature of the ghost has
subsequently appeared in the literature \citep{Jones04,Saunders04},
there has been no systematic discussion of its cause, expected
amplitude, and paths to mitigating the problem.  As it turns out, we
have identified two significant ways in which the ghost can arise.

Because the grating can recombine all of the light of the dispersed
spectrum that falls on the detector and reimage it into one resolution
element, the brightness of the ghost may be high relative to any
nearby spectral features, even if the efficiency of recombination by
the grating is in the range of $10^{-3}$.  Therefore, the deleterious
effects of the presence of this ghost are significant, especially for
multi-object spectroscopy, where each slit will produce a ghost that
may not be separated easily from the spectra of objects of interest.
Given the substantial efficiency advantages of VPH gratings, it is
important to understand the nature of this ghost, and how to use or
manufacture such gratings to mitigate or eliminate the effect.

In this paper, we present descriptions of the RSS and WIYN Bench
Spectrograph designs and present examples of the ghost.  We describe
the causes of the ghost and develop a model to predict the position of
the ghost and estimate its brightness.  Further, we suggest methods
for mitigating the effects of this ghost, for gratings already
designed as well as for future gratings.

\section{University of Wisconsin-Madison Spectrograph Developments}

Because the discovery and analysis of the VPH grating ghost was made
in the context of two specific spectroscopic instruments, we include
here a brief description of their capabilities. These salient
attributes contextualize the ghost discussion, and permit a more
general interpretation of our examples to other spectroscopic
systems. We begin with a basic discussion of spectrographic resolving
power, which frames the spectrographs' descriptions, as well as the
resulting impact on various modes to mitigate the ghost (\S 4.2).

By equating the size of the entrance slit with that of a spectral
resolution element through the appropriate series of transformations,
we get the following form for the resolving power:

\begin{equation}
\label{fullR}
R\equiv
\frac{\lambda}{\delta\lambda}
=
\frac{1}{w}
\frac{\partial w}{\partial\theta_s}
\frac{\partial\theta_s}{\partial\alpha}
\frac{\partial\alpha}{\partial\beta}
\frac{\partial\beta}{\partial\lambda}
\lambda
\end{equation}

\noindent where $\delta\lambda$ is the size of a resolution element in
wavelength; $w$ is the physical slit width; $\partial
w/\partial\theta_s$ is the inverse of the telescope scale, relating
the slit width to its angular width on the sky, $\theta_s$;
$\partial\theta_s/\partial\alpha$ is the angular magnification of the
collimator; $\partial\alpha/\partial\beta=1/r$, where $r$ is the
anamorphic magnification; and $\partial\beta/\partial\lambda$ is the
angular dispersion of the grating.  The angles $\alpha$ and $\beta$
are the incident and diffracted angles, respectively, at wavelength
$\lambda$.  For a spectrograph with fibers at the entrance rather than
a slit at the telescope focal plane, the second two terms would be
replaced with $\partial w/\partial\alpha= f_\mathrm{coll}$, where
$f_\mathrm{coll}$ is the collimator focal length.

For the standard plane-parallel, untilted fringe
VPH grating used in the Littrow configuration (i.e., $\alpha=\beta$)
this equation reduces to

\begin{equation}
\label{generalR}
R_L=\frac{f_\mathrm{coll}}{w}2\tan\alpha 
\end{equation}

\noindent In the case of a fiber-fed spectrograph like the WIYN Bench
Spectrograph, $w$ refers to the size of the fiber at the input to the
spectrograph.  For spectrographs with entrances at the telescope focal
plane, like RSS, the slit width relates to the image at the telescope
focal plane by

\begin{equation}
w=\theta_s f_\mathrm{coll}\frac{D}{d}
\end{equation}

\noindent where $d$ is the collimated beam diameter, and $D$ is the
primary mirror diameter.

\subsection{SALT/RSS}

The Robert Stobie Spectrograph for the Southern African Large
Telescope is a complex spectrograph with multiple operational modes
\citep{Kobulnicky03} that include long- and multi-slit spectroscopy
and spectropolarimetry.  It also contains a double-etalon Fabry-Perot
system, developed at Rutgers University \citep{Rangwala07}.

The fast beam speed of SALT ($F/4.2$) and prime focus position of RSS
posed great challenges that led to an all-refracting design,
incorporating VPH gratings \citep{Burgh03}.  The mechanical design
\citep{Smith06} of the spectrograph incorporates a camera articulation
mechanism and a grating rotation stage for ``on-the-fly''
repositioning of the camera and grating positions to take full
advantage of the tunable blaze properties of the VPH gratings.
Articulation angles as high as 100$\degr$, and thus incident grating
angles in the collimated beam, $\alpha$, as high as 50$\degr$, are
possible.  With telescope and collimated beam diameters of 11~m and
150~mm, respectively, and a $1\farcs2$ slit, a width well matched to
the median image size delivered at the focal plane of SALT, resolving
powers, following from Equation \ref{generalR}, as high as 5500 are
achieved.

The RSS grating complement includes five VPH gratings, ranging from
900 to 3000 lines~mm$^{-1}$, fabricated by Wasatch Photonics on
fused-silica substrates and one standard 300 lines~mm$^{-1}$ surface
relief grating from Richardson Gratings.  This complement allows for a
variety of resolutions and wavelength coverages to be obtained, with
maximum resolution achieved in the vicinity of several spectral
features of astrophysical importance, such as the \ion{Ca}{2} infrared
triplet ($\sim850$~nm), the H-$\alpha$ region ($\sim650$~nm), and the
H-$\beta$ and \ion{O}{3} region ($\sim500$~nm).

Multi-object observations with RSS are made using laser-cut carbon
fiber masks placed at the 8 arcminute field-of-view focal plane of the
telescope.  Up to 30 of these custom-cut masks can be installed in a
magazine on the spectrograph so that a wide range of potential science
programs are available for observation on any given night - well
suited to the intrinsically queue-scheduled nature of the telescope.

\subsection{WIYN Bench Spectrograph}

\begin{figure}
\epsscale{1}
\plotone{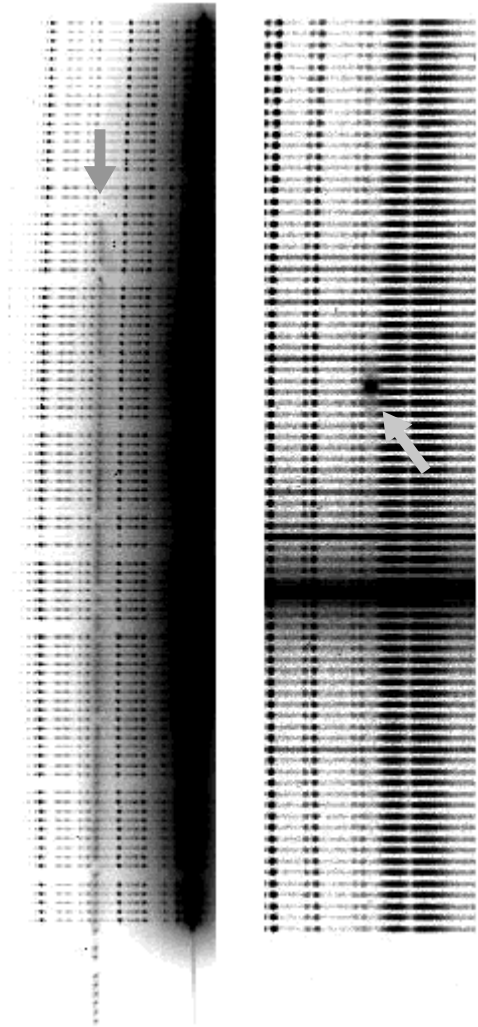}
\caption{Examples of WIYN Bench Spectrograph detection of the Littrow
ghost with a 740 lines~mm$^{-1}$ grating. Only a limited wavelength
range is shown.  Left: Ghosts generated by multiple fibers with
roughly equal brightness, using the red Hydra cable and the grating in
first order. Gaps, representing broken fibers, illustrate spatial
inversion of the ghost image.  The vertical displacement relative to
the direct spectrum arises from an out-of-plane misalignment of the
grating.  Right: Ghost generated from a bright source in a single
fiber, using the SparsePak IFU and the grating in second order.
Wavelength is left to right.
\label{fig2}}
\end{figure}

The WIYN Bench Spectrograph (Barden et al. 1993, 1994) is
bench-mounted, fiber-fed, and situated in a climate-shielded room two
stories below the telescope observing floor. Feeds for the 75~mm
spectrograph slit include two 100-fiber MOS cables (Hydra) with access to
a square degree on the sky, and two IFUs (DensePak and SparsePak)
covering 0.3-1~arcmin field of view. Fibers are 200--500~$\mu$m in
diameter, or roughly $1\farcs8$ to $4\farcs6$ at the $F/6.3$ Nasmyth
focus of the telescope (8.8 arcsec~mm$^{-1}$). Performance features of
the system with SparsePak are presented in \citet{Bershady05}.

The spectrograph consists of an on-axis parabolic reflecting
collimator ($f_\mathrm{coll}=1021$~mm), one or two grating turrets,
and an all-refractive camera\footnote{A second, catadioptric camera
can be used for low-dispersion work into the blue, but is lossier
because of a central obstruction filled in by fiber focal-ratio
degradation.}. The spectrograph can be optimized for a wide range of
gratings (echelle and low order surface-relief gratings, as well as
VPH) because of its adjustable camera-collimator, grating and CCD
focal-plane angles, as well as adjustable grating-camera distance.
Double-turret configurations allow for a fold-flat to accommodate
small grating angles, or a second grating. In contrast to RSS, only
the grating angle of the primary is currently remotely controllable,
with the remaining degrees of freedom requiring manual adjustment.

In Littrow, single-grating configurations, the Bench achieves $R =
17,800$ at $\alpha = 50\degr$ for a $1\farcs2$ effective slit width.
At comparable grasp ($A\Omega$) with RSS on SALT (scaling from
respective clear apertures of 60 and 8~m$^2$), the Bench achieves $R =
6,500$ ($3\farcs3$ effective slit width), or 15\% higher than
RSS. However, the Bench can be used at higher angles, with VPH
gratings optimized for $\alpha$ as large as 70$\degr$ now being
implemented; in non-Littrow configurations (e.g., a
316~lines~mm$^{-1}$ echelle blazed at 63$\fdg$4) yielding anamorphic
demagnification factors that boost $R$ by factors of 40-50\%; and in
double-grating configurations where one or both gratings have
transmissive diffraction.  Reported here are the results of a
740~lines~mm$^{-1}$ VPH grating developed by Sam Barden with CSL
(Centre Spatiale de Liege), made on float-glass and post-polished to
60\% Strehl at Lawrence Livermore Labs via NOAO contract, and
implemented on the Bench at low angles of $17-24\degr$ via a
double-turret configuration using a fold-flat in the primary turret.

\section{Ghost Model}

\begin{figure*}[t]
\epsscale{1}
\plotone{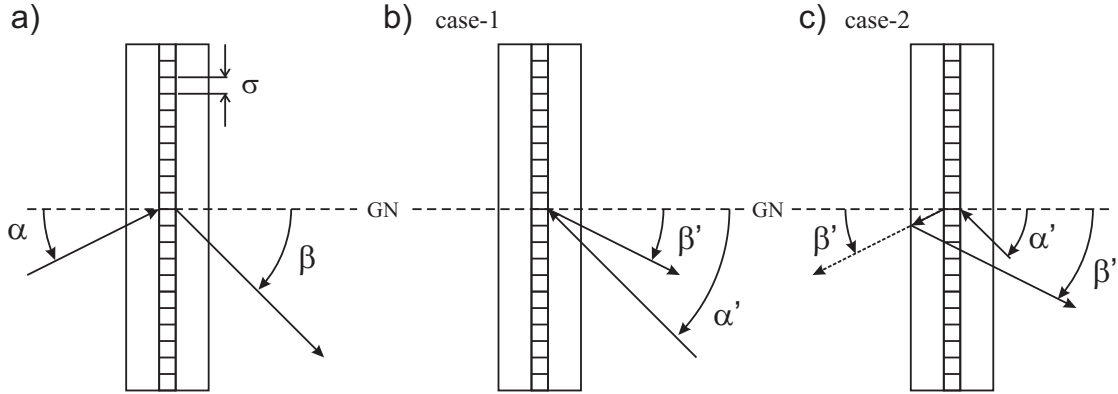}
\caption{Left (a): In-plane (i.e. $\gamma=0$) geometry for first
(transmissive) pass through a plane-parallel VPH grating (not to
scale).  Middle (b): Geometry for case-1 ghost: reflection off of the
grating ($\beta'$) after reflection off of the CCD and recollimation
by the camera ($\alpha'$). Right (c): Geometry for case-2 ghost:
transmission through the grating ($\alpha'$), again after CCD
reflection and camera recollimation, then reflection off the substrate
air-glass interface, and finally zeroth order grating transmission
($\beta'$). Angles are measured relative to the grating normal (GN).
Note: The refractive changes of angles at the substrate interfaces
have been ignored (see text).
\label{fig3}}
\end{figure*}

Based on the appearance of ghost images in two independent
spectrographic systems, we have constructed a physical model that
allows us to reproduce their behavior and predict a more general ghost
phenomenon.  The incident light is dispersed by the transmission
grating and is focussed onto the CCD by the camera.  A sizable
fraction of the light, roughly 10\% (perhaps even higher at
wavelengths where the QE is low), is reflected from the surface of the
CCD and recollimated by the camera. Upon reaching the grating, it is
recombined and reflected back through the camera on a third pass.
Depending on the order of interaction with the grating, this may
result in an image of the spectrograph entrance focal plane or another
spectrum with (possibly) different dispersion on the detector plane.

There are two paths for this recombination: (1) dispersive (i.e.,
non-specular) reflection off of the grating back toward the camera
(see Figure \ref{fig3}b; we refer to this as case-1 or ``reflective''
ghost); or (2) dispersive transmission through the grating, then
reflection off of the air-glass interface of the grating substrate on
the far side, followed by a zeroth order transmission back through the
grating, sending the light in the direction of the camera (See Figure
\ref{fig3}c; we refer to this as case-2 or ``transmissive'' ghost).
Both recombination paths produce essentially the same effect for
plane-parallel gratings substrates and untilted fringes, though in
general they have different efficiencies, and do not share the same
set of solutions (\S 4).

To understand how these ghost mechanisms work in quantitative detail, we
start with the generalized grating equation, given by:

\begin{equation}
\label{gratingequation}
m\lambda=n_i\sigma(\sin\alpha_i+\sin\beta_i)\cos\gamma_i
\end{equation}

\noindent where $n_i$ is the index of refraction of the
medium\footnote{In this work we will ignore any wavelength dependence
of the index of refraction.}, $\alpha_i$ is the incident angle of the
light relative to the grating normal in the plane perpendicular to the
grating grooves, $\beta_i$ is the diffracted angle for order $m$,
$\sigma$ is the groove spacing, and $\gamma_i$ is the incident angle
in the plane parallel to the grooves. This holds for passage through
the DCG ($i=2$), the substrate material ($i=1$), or the air ($i=0$).
For the sake of simplicity we will, unless specified otherwise, refer
to the angles in air and use the angles without subscripts,
i.e., $\alpha=\alpha_0$, etc.  See Figure \ref{fig3}a for a schematic
representation of this geometry.

On the light's first pass through the grating, the output angle is then:

\begin{equation}
\label{betadirect}
\sin\beta=\frac{m\lambda}{\sigma\cos\gamma}-\sin\alpha
\end{equation}

\noindent The dependence of $\beta$ on $\gamma$ is what is responsible
for spectral line curvature, as seen in Figures 1 and 2.

After reflection from the detector surface and subsequent
recollimation by the camera, the light interacts with the grating again,
with an output angle following:

\begin{equation}
\label{betaprime}
\sin\beta'=\frac{m'\lambda}{\sigma\cos\gamma'}-\sin\alpha'
\end{equation}

\noindent 
Note that for generality we have allowed the second diffraction to be
in another order.  For the case in which the surface of the grating
substrate is parallel to the grating, the output angle is the same
after reflection from the substrate air-glass interface and thus this
equation holds for both the reflective and transmissive recombinations
(see Figure \ref{fig3}b and \ref{fig3}c respectively).

Because the reflection off the detector happens at a focus, where the
position on the detector is conjugate with angle in the collimated
beam, the angle that the light makes as it is recollimated will be the
same as it was after dispersion.  Therefore, the incident angle for
the second encounter with the grating is equal to the diffracted angle
after the first, i.e., $\alpha'=\beta$, and thus:

\begin{equation}
\sin\beta'=\frac{m'\lambda}{\sigma\cos\gamma'} -
\left[\frac{m\lambda}{\sigma\cos\gamma}-\sin\alpha\right]
\end{equation}

\noindent Because $\cos\gamma'=\cos\gamma$ this reduces to

\begin{equation}
\label{ghosteq}
\sin\beta'=\frac{\Delta m\lambda}{\sigma\cos\gamma}+\sin\alpha
\end{equation}

\noindent where $\Delta m\equiv(m'-m)$.  The angular dispersion of the
ghost will be:

\begin{equation}
\label{dispersion}
A'=\frac{\partial\beta'}{\partial\lambda} = \frac{\Delta
m}{\sigma\cos\beta'\cos\gamma}
\end{equation}

\noindent Because the angular dispersion of the direct spectrum is $A
= m/(\sigma\cos\beta\cos\gamma)$ then the relative dispersions between
the direct and the ghost spectra for $\beta'=\beta$ will be

\begin{equation}
\label{reldispersion}
\frac{A'}{A}=\frac{\Delta m}{m}
\end{equation}

It follows from Equations \ref{ghosteq} and \ref{reldispersion}
that an important parameter for determining the position and
dispersion of the ghost is $\Delta m$, the relative change in order of
the ghost to the direct spectrum.  A few interesting cases are
presented in the following sections.

\subsection{Ghost Modes}
\subsubsection{Narcissistic Ghost ($m'=0$)\label{narcissistic}}

The trivial case where $m'=0$ produces a ``narcissistic'' ghost.  In
this case, the grating acts as a mirror, i.e., $\beta'=-\alpha'$ and
the ratio of the dispersions will be $A'/A=-1$, resulting in an
inverted spectrum.  In the case of VPH gratings, it may be likely that
a simple reflection off of the air-glass interface of the camera-side
substrate produces this ghost with more efficiency than a zeroth-order
reflection from the DCG.  This case is particularly harmful when
$\beta=0$ \citep{McCandliss98} or when $\beta$ is less than the
viewing angle of the camera.

\subsubsection{Littrow Ghost ($\Delta m=0$\label{litghost})}

If the recombination by the grating is in the same order as the initial
diffraction then $\Delta m=0$.  Equation \ref{ghosteq} is then simply
$\beta'=\alpha$, and the light follows the path of the Littrow
wavelength, independently of wavelength, i.e., the light is fully
recombined.  This results in an image of the spectrograph entrance
slit(s), \textit{without line curvature}, on the detector at the
location of the Littrow wavelength, as illustrated in Figures
\ref{fig1} and \ref{fig2}.

VPH gratings are most efficient near the Bragg wavelength, i.e., when
the light is ``reflected'' from the plane of the grating fringes:

\begin{equation}
\label{braggreflection}
\alpha_2 - \phi = \beta_2+\phi
\end{equation}

\noindent where $\alpha_2$ and $\beta_2$ are the angles of incidence
and reflection, respectively, inside the DCG and $\phi$ is the tilt of
the fringes relative to the grating normal (see Figure \ref{fig5}).
For a grating with untilted fringes\footnote{The general case, which
includes tilted fringes, is discussed in \S \ref{futuregratings}.},
i.e., $\phi=0$, like the ones built for RSS and the WIYN Bench
Spectrograph, this results in the highest efficiencies being produced
at the Littrow condition, i.e., $\alpha_2=\beta_2$.  For a
plane-parallel grating, the DCG is sandwiched between flat substrates,
and it also holds that $\alpha=\beta$.  Thus, the standard operating
mode is in this Littrow configuration and the central wavelength on
the detector is the ``Littrow wavelength'', defined as

\begin{equation}
\label{littwave}
\lambda_L=\frac{2\sigma\sin\alpha}{m}.
\end{equation}

\noindent Because this is the standard configuration, VPH spectrographs are
particularly sensitive to the Littrow ghost.

Since the camera-collimator angle, $\Phi$, is equal to
$\alpha+\beta'$ for the ghost, $\Phi=2\alpha$ and
$d\Phi/d\alpha=2$, i.e., when the grating is moved by $\delta\alpha$
away from a Littrow configuration, the ghost moves twice that angle,
consistent with what was observed\footnote{An interesting side note is
that this sensitivity of the ghost position can be utilized as a
calibration of the grating rotation angle.  For RSS, the ghost moves
one unbinned pixel per 4.5 arcseconds of grating rotation.}.

\subsubsection{Other recombinations}

In the case that $\Delta m\ne0$, the light will not fully recombine
and the ghost will take the form of a spectrum. The zeroth order,
$m'=0$, results in the ``narcissistic'' ghost described above, but for
gratings that operate at second order or higher there exists the
possibility of other ghosts.

In general, the configurations that will result in the production of
this ghost are ones where the ghost is in a Littrow configuration, and
thus the recombined light follows the direction of the primary
spectrum, i.e., $\sin\beta'=\sin\beta$. Combining Equations
\ref{betadirect} and \ref{ghosteq} produces the wavelengths at which
this will occur:

\begin{equation}
\label{lamghosts}
\lambda = \frac{2(\sigma/m)\sin\alpha}{1-\Delta m/m} =
\frac{\lambda_L}{1-\Delta m/m}
\end{equation}

\noindent where $\lambda_L$ is the Littrow wavelength (Equation
\ref{littwave}) for the given grating's line ruling, angle and order
of primary use.  Solutions only exist for $\Delta m/m<1$, therefore
$m'<2m$.\footnote{In principle, the order for the ghost is further
constrained by the fact that the wavelength must be diffracted by an
angle less than 90$\degr$ for the first pass through the grating, so
$\sin\beta<1$.  The result is that
\[m'<\frac{2m}{1+\sin\alpha}.\]
\noindent In practice it will be even more constrained by the fact
that the wavelength must fall on the detector to be reflected in the
first place, i.e., $\beta_c+\delta>\beta>\beta_c-\delta$, where
$\beta_c$ is the angle of the central wavelength and $\delta$ is the
camera half-angle field-of-view.}  Thus, only gratings in second order
or higher can see this ``cross-order'' ghost.  What is observed is a
partially recombined ghost spectrum ($\Delta m/m < 1$), which may be
inverted in wavelength ($\Delta m<0$) or not ($\Delta m>0$) as per
Equation \ref{reldispersion}, having the above wavelength in common
with the direct spectrum.

In practice, this type of ghost is of relevance to primary spectra
produced in off-Littrow configurations that may include light at
significant power in the Littrow wavelength for another order.  While
existing VPH gratings are designed to work in Littrow, they can be
used off-Littrow.  Future gratings with tilted fringes (\S 4.2.1) will
also operate in non-Littrow configurations where this ghost may arise
-- if used in second or higher order.  Furthermore, a ``cross-order''
ghost was observed in the Gemini Near-Infrared Spectrometer (GNIRS)
\citep{Joyce03} using surface-relief reflection echellettes,
demonstrating that our general ghost model is not intrinsic to VPH
gratings nor to the use of a primary Littrow configuration.

In the GNIRS case, a 110.5 lines~mm$^{-1}$ surface-relief reflection
grating, blazed for 6.79~$\mu$m in first-order Littrow, was used at
$\alpha=39\fdg1$ and $\beta=12\fdg1$ in second order ($m=2$). This
yields a central wavelength of 3.8~$\mu$m.  A Littrow configuration
for the ghost, $\beta'=\beta=12\fdg1$, also occurs at 3.8~$\mu$m for a
first-order reflection ($m'=1$), following directly from Equation
\ref{lamghosts} (with $\lambda_L=5.7\mu$m for this grating at $m=2$
and $\alpha=39\fdg1$).  The result is fully consistent with what they
noted: the ghost's resolving power was half the primary spectrum, and
the length of the ghost was exactly half of the detector width.  These
effects arise because the relative change in dispersion will be,
according to Equation \ref{reldispersion}, $A'/A=-1/2$, and only the
wavelengths in the primary spectrum will contribute.  Should the
grating have significant efficiency across multiple orders for a given
wavelength, more ghosts, at wavelengths satisfying Equation
\ref{lamghosts}, may be observable. Indeed, this is the case for the
GNIRS grating suite, and \citet{Joyce03} mentions having observed
ghosts in other configurations.

\subsection{Off-Axis Ghosts and Multi-Object Spectroscopy}

For spectrographs that employ a single entrance slit, or have a fiber
bundle aligned as a long slit, like the WIYN Bench Spectrograph, the
Littrow ghost will be constrained to a single area on the detector
(see left panel of Figure \ref{fig1} or Figure \ref{fig2}). However, for a
multi-object spectrograph like RSS, which uses multiple apertures at
the focal plane, there arise ghosts from the spectra through
each aperture.

For light that arrives at the grating off-axis\footnote{An object that
is off-axis by an angle $x$ in the focal plane field-of-view will have
an optical magnification equal to the ratio of the focal lengths of
the telescope and spectrograph collimator and arrive at the grating
off-axis by $\Delta\alpha = x f_\mathrm{tel}/f_\mathrm{coll}$.} in the
spectral dimension by an amount $\Delta\alpha$, the diffracted angle
is

\begin{equation}
\sin\beta=\frac{m\lambda}{\sigma\cos\gamma}-\sin(\alpha+\Delta\alpha)
\end{equation}
 
\noindent and following the same steps as for the on-axis case
(Equations \ref{betadirect} and \ref{betaprime}, i.e., a reflection)
results in a ghost angle of

\begin{equation}
\sin\beta'=\frac{\Delta m\lambda}{\sigma\cos\gamma}+\sin(\alpha+\Delta\alpha)
\end{equation} 

\noindent For the Littrow ghost, $\Delta m$=0, and

\begin{equation}
\beta'_L=\alpha+\Delta\alpha
\end{equation}

\noindent resulting in a ghost positioned opposite the center of field
from its Littrow wavelength in the primary spectrum (the mirroring is
in both dimensions) .  Each aperture in the slitmask will have such a
ghost (see right panel of Figure \ref{fig1}) and thus the overall
effect is an image of the focal plane on the detector mirrored through
the optical axis.

\subsection{Recombination-Ghost Efficiency of VPH Gratings}

Here we focus on the fully recombined $\Delta m = 0$ Littrow ghost,
but our development is general in the context of the integrated ghost
flux.  The detected integrated brightness of the ghost will be:

\begin{equation}
\label{ghostbrightness}
B=\int_{\lambda_1}^{\lambda_2} F(\lambda) R_\mathrm{CCD}(\lambda) 
T_\mathrm{cam}^2(\lambda) \epsilon_{m'}(\lambda)  d\lambda
\end{equation}

\noindent where $F(\lambda)$ is the impinging flux of the direct
dispersed spectrum on the CCD; $R_\mathrm{CCD}$ is the CCD
reflectivity; $T_\mathrm{cam}$ is the throughput of the camera,
including the reflective losses at the camera-side grating-substrate
air-glass interface (squared because there are two passes);
$\epsilon_{m'}$ is the recombination efficiency of the grating in
order $m'$, and $\lambda_1$ and $\lambda_2$ are the lower and upper
wavelengths that impinge on the CCD(s) in the primary spectrum.

In the case-1 reflective recombination ghost, $\epsilon_{m'}$ will be
$\epsilon^{R}_{m'}$, the reflective diffraction efficiency in order
$m'$. In the case of the transmissive recombination, the efficiency
is:

\begin{equation}
\label{transeff}
\epsilon_{m'}=\epsilon^{T}_{m'}\times R_\mathrm{sub}\times \epsilon^T_0
\end{equation}

\noindent where $\epsilon^T_{m'}$ is the transmission efficiency in
order $m'$, $R_\mathrm{sub}$ is the reflectivity of the substrate
air-glass interface, and $\epsilon^T_0$ is the zeroth order
transmission efficiency of the grating. 

In the extreme case where $R_\mathrm{CCD}=1$ and $T_\mathrm{cam}=1$,
we would expect the lower limit for the efficiency, $\epsilon_{m'}$,
of the grating recombination to be of the same order as the ratio of
the integrated flux in the ghost to the integrated flux of the direct
spectrum. For RSS and the Bench, this ratio is typically observed to be a
few times $10^{-4}$.  With CCD reflectivities in the 10\% range and
more realistic camera throughputs, the efficiency, $\epsilon_{m'}$,
would be at least a few times $10^{-3}$.

\begin{figure}[t]
\epsscale{1}
\plotone{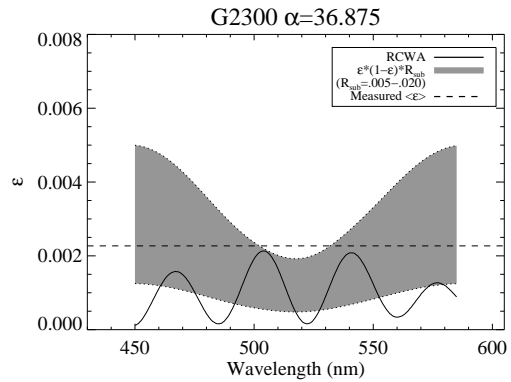}
\caption{Efficiencies of reflective and transmissive recombination for
a specific RSS grating configuration, defined in text. The solid line
is the RCWA prediction for the first-order reflection efficiency.  The
shaded region shows the predicted efficiency of the transmissive
ghost, with lower and upper bounds on the substrate reflectivity of
0.5\%--2\%.  Note: reflective losses from the camera-side substrate
air-glass interface, which will be the same for both cases, are not
included.  The dashed line is the measured average value of the
recombination efficiency for this grating configuration as derived from
$\epsilon=(B/F_\mathrm{d})/(T_\mathrm{cam}^2*R_\mathrm{CCD})$, which
follows from Equation \ref{ghostbrightness}, where $B$ and
$F_\mathrm{d}$ are the integrated fluxes of the ghost and direct
spectrum respectively, $T_\mathrm{cam}$ is the measured camera
throughput of RSS, and where we have assumed that $R_\mathrm{CCD}$ is
$1-QE_\mathrm{CCD}$, i.e., no absorption.
\label{fig4}}
\end{figure}

Figure \ref{fig4} shows the efficiencies of the transmissive and
reflective recombinations, estimated for one particular RSS grating
configuration (2300 lines~mm$^{-1}$ grating used at $\alpha\sim37\degr$)
using a rigorous coupled-wave analysis code (RCWA)\footnote{RCWA code,
written in C, was generously provided by Gary Bernstein, who
implemented the methods of \citet{Moharam83}.}. We show a
range of efficiencies for the transmissive recombination (case-2),
corresponding to substrate air-glass interface reflectivities of
0.5\%-2\%, and assuming that the zeroth order transmission is equal to
one minus the first order transmission ($\epsilon^T_0=1-\epsilon^T_1$)
-- an assumption supported by the RCWA results.  For this
recombination case, $\epsilon$ is in the few times $10^{-3}$ range,
consistent with that observed, and factors of a few more efficient
than the reflective recombination (case-1) ghost.  

The brightness of the case-2 recombination will depend on the
performance of the anti-reflection coating, which, in turn, will
depend on the particular coating recipe and incident angle.  In most
cases, we believe the reflectivities should be in our adopted range,
and though the case-2 ghost may be the brighter one in general, we do
expect that both ghost production methods will contribute to the
overall intensity of the ghost.  If, however, the grating is designed
for use at very large incidence angles ($\alpha>50\degr$), such as
used on the Bench, the case-2 ghost could be significantly stronger
than the case-1 ghost.  Because of their different response to
mitigation (\S 4), it is important to track both cases.

For the $\Delta m=0$ ghost, there is complete recombination and the
relative brightness of the ghost to the direct spectrum is enhanced by
the number of resolution elements on the detector, which in modern
spectrographs can be as high as $10^3$ (see Equation \ref{nresel}).
Therefore, the flux of the ghost may be a significant fraction of the
flux per resolution element of the direct spectrum, and, depending on
the character of the direct spectrum and the placement of the ghost,
the ghost may actually be brighter than its surroundings.  This is
seen in Figures \ref{fig1} and \ref{fig2}.

\section{Ghost Mitigation}

The discovery of this ghost was unexpected.  Through testing and
on-sky observations it has become clear that it can be bright enough
to cause significant disruption of the primary data.  In the following
sections we describe potential methods for minimizing the
effects of the ghost for both existing and future spectrographs.

\subsection{Existing gratings}

In spectrographs that have already been designed and built to use VPH
gratings there are several options to limit the effects of the ghost. 

\subsubsection{Off-Littrow configurations\label{off-littrow}}

As mentioned in Section \ref{litghost}, the ghost moves when the
grating angle is changed.  However, for a fixed camera-collimator
angle, $\Phi=\alpha+\beta$, the central wavelength on the detector, to
zeroth order, does not move with grating tilt.  This follows from
\begin{equation}
\label{dPhidalpha}
\frac{d\Phi}{d\alpha} = 1+\frac{\partial\beta}{\partial\alpha} =
1-\frac{\cos\alpha}{\cos\beta}
\end{equation}
and the fact that near Littrow $\alpha\approx\beta$, so
$d\Phi/d\alpha\approx0$. 

Consequently, the position of the ghost can be moved with small
movements of grating angle, and subsequent small changes in central
wavelength and dispersion. For a single-slit spectrograph, to move the
ghost completely off the detector requires a grating rotation of
$\Delta\alpha=\delta/2$, where $\delta$ is the camera half-angle field
of view in the dispersion direction\footnote{Camera FOV = $2\delta =
\arctan(d/f_\mathrm{cam})$, where $d$ is the detector size in the
dispersion direction and $f_\mathrm{cam}$ is the camera focal
length.}.  In the case of detectors with CCD mosaics, like the
three-chip RSS detector, the gap between chips may provide a
convenient place to put the ghost, which can be accomplished with a
smaller grating rotation.  For a multi-object spectrograph, the
rotation will need to be larger, to accommodate the off-axis ghosts as
well.

The downside to such a maneuver is that the VPH blaze efficiency is
shifted when operated off-Littrow and the change in efficiency may be
significant even for small moves such as that needed to put the ghost
in the detector gap for a single-slit spectrograph.  For example, the
RSS camera has a half-angle FOV of $\delta\sim8\degr$, requiring a
grating rotation of $\Delta\alpha=\delta/2=4\degr$ to remove the ghost
completely ($\Delta\alpha=2.4\degr$ for the Bench); however, it is a
3-chip mosaic, so to put the ghost in the detector gap would only
require a move of $\Delta\alpha=(\delta/3)/2=1\fdg33$.  For a typical
RSS or Bench grating configuration, going off-Littrow by this amount
produces a blaze-shift that can reduce the efficiency at one end of
the spectrum by as much as $\sim$20\% (with a corresponding increase
at the other end).  In practice, the details of the blaze-shift will
depend on the grating and the grating angle.

In multi-object mode the grating would have to be tilted even farther
to accommodate the ghosts generated by off-axis slits.  For RSS as
much as an additional $2\fdg5$ of grating rotation could be necessary
to remove all ghosts from the detector, depending on the distribution
of slitlets on the multi-object slitmask.

\subsubsection{Out-of-plane configurations}
Another option would be to have the grating placed out-of-plane, i.e.,
$\gamma\ne0$ for on-axis light.  To move the ghost completely off the
detector may then require a smaller angle out-of-plane than the
in-plane, off-Littrow configuration if the detector is wider in the
dispersion direction.  For example, the RSS detector array has a 1.5
aspect ratio, so the half-angle field of view in the spatial direction
is 2/3 the value of the spectral direction, or about $5\fdg33$.  The
corresponding value for the Bench, given the 2:1 aspect ratio of the
used portion of the CCD, is $2\fdg4$. An out-of-plane tilt of the
grating of this amount would move the entire ghost off the detector,
for both longslit and multi-object modes.  For a given $\beta$, there
may be no resultant blaze shift; however, the center of the line
curvature will shift, resulting in a potentially substantial
enhancement of line-curvature on one side of the spectral lines.  The
practicality of implementing this configuration will be spectrograph
dependent.  For example, it is relatively straightforward to modify
the grating mounts on the Bench, but strong mechanical constraints
prohibit this on RSS.

\clearpage
\subsubsection{Dithering}
In practice, we postulate that the best method for mitigating the
effects of the ghost is procedural, achieved by employing a
``dithering'' procedure, in which an observation is split into two (or
more) exposures, each with a different instrument physical
configuration.  There are three potential spectrograph dither types:
the camera-collimator angle, $\Phi$, is kept the same, but the grating
angle, $\alpha$, is changed to an off-Littrow configuration; $\alpha$
is kept the same, but $\Phi$ changed; or both angles are changed such
that a Littrow configuration is maintained.  An additional form of
dithering would be to maintain a fixed spectrograph configuration but
nod the telescope so that the object of interest occupies a different
position along the slit.

\begin{enumerate}
\item Grating rotation only:

In this case, the position of the ghost will move by twice the grating
angle change, as per Section \ref{litghost}, with little to no change
in position of the primary spectrum (see Equation \ref{dPhidalpha}).
However, the dispersion of the primary spectrum will change somewhat
by such a grating rotation, making it difficult to simply co-add the
two exposures and requiring independent wavelength and/or flat-field
calibrations.  As mentioned above, the move to an off-Littrow
configuration will shift the blaze; however, if the goal is to only
move the ghost by a resolution element or two, the shift is
negligible.  This option may work best for spectrographs like the WIYN
Bench Spectrograph, for which the camera-collimator angle may not be
changed during the course of an observation, but grating angle can.

\item Camera-collimator angle change only:

A change in $\Phi$ may be desirable in a CCD mosaic in order to
recover any wavelength coverage lost in the detector gaps.  If this
were done, with fixed grating angle, the central wavelength would be
different. However, with no change in $\alpha$, the Littrow ghost
will continue to follow the path of the Littrow wavelength, i.e., the
ghost shifts the exact same amount as the direct spectrum, and no
separation between the two is achieved. In short, this approach
does not work.

\item Maintain Littrow configuration:

Perhaps the best solution for spectrographs like RSS, which allow for
remote control of both $\alpha$ and $\Phi$, is to adjust both while
maintaining a Littrow configuration.  This results in movement of the
direct spectrum, because of the change in $\Phi$, so that the
wavelengths that fall in the detector gaps are recovered.  But it also
results in the movement of the ghost because of the change in
$\alpha$. However, as in option (1) above, additional calibration and
wavelength solutions are needed, and no simple co-addition of the data
is possible.

\item Telescope Nod:

While maintaining a fixed spectrograph configuration, a telescope nod
can be performed to place the object of interest at different spatial
positions along the slit. This either moves the ghost relative to the
source (source-limited regime; see Figure \ref{fig2}, right panel), or
moves the source to intersect the ghost at a different wavelength
(background-limited regime; see Figures 1 or \ref{fig2}, left
panels). In general, the feasibility of this solution depends on the
spatial extent of the source, and the degree of line-curvature in the
spectrum. The advantage of this approach is that it requires no
additional instrument setup or calibration.

\end{enumerate}

Each of the three methods (1), (3) and (4), will mitigate the effect
of the ghost overlapping an area of interest in the direct spectrum;
however, the extent to which each is desirable depends on the specific
science goals of the observation and the technical limitations of the
telescope/spectrograph system (i.e., the cost and feasibility of
multiple calibrations). If continuous wavelength coverage is of high
importance for multi-detector systems such as RSS (where detector gaps
are present), then methods (1) and (3) would be best. They require the
additional overhead associated with both the reconfiguration of the
spectrograph as well as the need to recalibrate the reconfigured
system.  Otherwise, method (4) is likely the most natural solution
because it is often the case that one wishes to dither a source along
the slit (to minimize or average slit or detector variations and
defects).

Since method (4) has some clear operational advantages, we elaborate
under what conditions it will work.  In all cases when the ghost is
dominated by flux from the source, and the source is small relative to
the slit-length, the ghost's discrete image in the spatial dimension
(along the slit) is reflected about the center of the detector
(barring any out-of-plane grating misalignments). Consequently, a
maneuver to position the source off-axis will move the ghost by an
equal amount in the opposite direction (see right-hand panel of Figure
\ref{fig2}, where the ghost is offset from its direct spectrum).  For
point-source objects, where the spectrum (and hence ghost) is
source-limited, small nods would suffice, with little to no impact on
the wavelength coverage, dispersion or line curvature.

For background-limited observations, the ghost is generated
predominately by the sky lines and thus is uniform in intensity at all
slit positions (corrected for vignetting). Small nods do not move the
direct spectrum away from the ghost. However, at high enough grating
angle, one may take advantage of the line curvature to place the
source at two slit locations that have the ghost intersection with the
direct spectrum at different wavelengths. Line curvature follows from
differentiating Equation \ref{betadirect} with respect to $\gamma$:

\begin{equation}
\cos\beta\frac{\partial\beta}{\partial\gamma} =
\frac{m\lambda}{\sigma}\frac{\tan\gamma}{\cos\gamma}
\end{equation}

\noindent At the Littrow wavelength (Equation \ref{littwave}) and for
small values of $\gamma$, this reduces to:

\begin{equation}
\frac{\partial\beta}{\partial\gamma}=2\gamma\tan\alpha
\end{equation}

\noindent which gives

\begin{equation}
\label{linecurve}
\beta_\gamma=\gamma^2\tan\alpha
\end{equation}

\noindent where $\beta_\gamma$ is the change in $\beta$ from that for
the Littrow wavelength on the optical axis, $\gamma=0$.  The spectral
lines are curved parabolically and shift to higher angles, i.e.,
longer wavelengths, along the slit.

The minimum movement necessary would be one that changes the
diffracted angle by an amount equal to that for a resolution element,
which in terms of $\beta$ follows from the generalized resolving power
equation (Equation \ref{fullR}) and can be written as

\begin{equation}
\delta\beta = w\frac{\partial\theta_s}{\partial
w}\frac{\partial\alpha}{\partial\theta_s}\frac{\partial\beta}{\partial\alpha}
= \frac{wr}{f_\mathrm{coll}}
\end{equation}

\noindent At Littrow, the anamorphic factor, $r$, equals one.  Setting
the above equation equal to Equation \ref{linecurve} and solving for
$\gamma$ produces

\begin{equation}
\gamma = \sqrt{\frac{w}{f_\mathrm{coll}\tan\alpha}}
\end{equation}

\noindent which, in terms of the filled-slit resolving power at
Littrow (Equation \ref{generalR}) can be written as

\begin{equation}
\gamma=\sqrt{\frac{2}{R_L}}
\end{equation}

\noindent This angle is measured in the collimated beam and can be
related to the on-sky angle in the spatial direction, $\theta$, by
$\gamma=\theta f_\mathrm{tel}/f_\mathrm{coll}$.  For RSS at the
highest resolving powers, (i.e. $R_L=5500$ for $\alpha=50\degr$ and
$\theta_s=1\farcs2$), this amounts to a minimum of $\sim1\arcmin$ nod.
For the lowest resolving power settings ($R_L\sim1000$) the nod would
be about twice as large.

\subsubsection{VPH Grisms}
It should be noted that a ghost should appear in VPH grism
spectrographs as well; however, we expect only the reflective
recombination ghost to contribute.  In the transmissive recombination
case the light would not be redirected into the camera because of the
typically large prism angles in a grism. 

The Low Resolution Spectrograph for the Telescopio Nazionale Galileo
has reported observing a ghost while commissioning new VPH grisms in
late 2005\footnote{See
\url{http://www.tng.iac.es/news/2005/12/13/lrs-vph/}}.  A ghost has
also been seen in the new grism for the Andalucia Faint Object
Spectrograph and Camera on the Nordic Optical Telescope\footnote{See
\url{http://www.not.iac.es/instruments/alfosc/grisms/grism17.html}}.
As reported, these ghosts match well the model discussed in this work.

In the case of grism spectrographs, where zero-deviation
configurations are implemented, it is not possible to change the grism
configuration between exposures.  One would have to move the position
of the object on the telescope focal plane, either by nodding the
telescope along the slit-axis, as mentioned in the previous section,
or by repositioning the slit in the dispersion direction (with
accompanying shift in central wavelength), which some grism
spectrograph designs can accommodate.  Because the ghost is reflected
about the optical axis, the former case would separate the object and
ghost in the spatial direction and the latter in the spectral
direction.

\subsection{Future gratings\label{futuregratings}}

\begin{figure}[t]
\epsscale{1}
\plotone{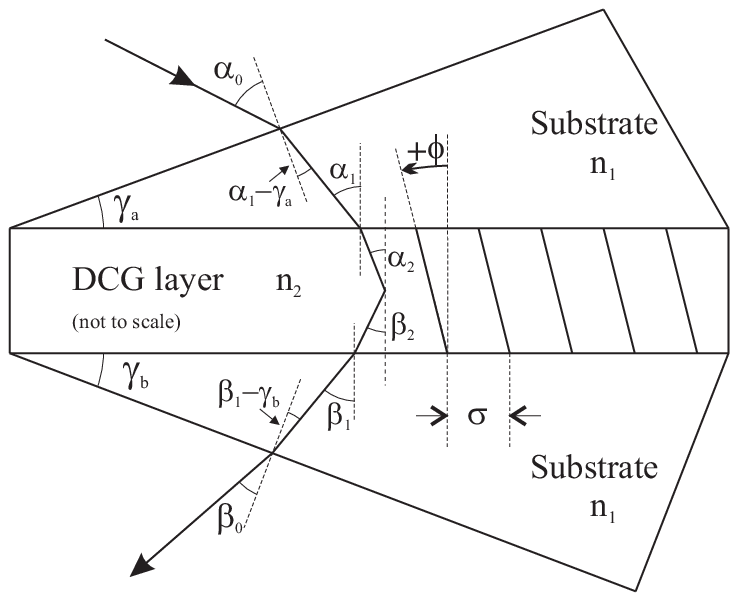}
\caption{General geometry for prism-immersed VPH grating with tilted
fringes. For tilted fringes only, $\gamma_a=\gamma_b=0$.  For prism
immersion only, $\phi=0$. In general, the two substrate refractive
indices can be different.
\label{fig5}}
\end{figure}

Given the freedom to design a VPH spectrograph with the foreknowledge
of the Littrow ghost, there are ways to mitigate the problem by
modifying the grating design.  The possibilities include tilting the
fringes and/or applying a wedge to the grating substrates.  The
concept of fringe-tilting in plane-parallel VPH gratings to remove the
Littrow ghost is mentioned in \citet{Saunders04}, while
\citet{Baldry04} describe prismatic substrates for VPH gratings,
though not in the context of ghost removal.  Figure \ref{fig5} shows
the geometry for a generalized VPH grating with both tilted fringes
and wedged substrates.  We consider only cases in which the two
substrates have the same index of refraction.

In addition to mitigating the ghost problem, redesigning the grating
can have an impact on the resolving power and the number of resolution
elements, and so we consider these two information metrics in tandem.
For this discussion, we refer back to Equation \ref{fullR}, the
generalized form for the resolving power.  The first three terms,
$(1/w)\times\partial w/\partial\theta_s \times
\partial\theta_s/\partial\alpha$, $(= f_\mathrm{coll}/w)$, represent
the fixed geometry for a given spectrograph collimator and slit width,
and remain constant for all grating configurations.  The final three
terms depend on the geometry of the grating setup, however, and we
define this as the resolution merit function $(1/r)\times
d\beta/d\mathrm{log}\lambda$, following \citet{Bershady07}.

Evaluating at the Bragg wavelength, this merit function takes the
following form:

\begin{equation}
\frac{1}{r}\frac{d\beta}{d\log\lambda} =
\frac{\cos(\alpha_1-\gamma_a)}{\cos\alpha_1}
\frac{\sin(\alpha_2-\phi)}{\cos\alpha_0}2n_2\cos\phi
\end{equation}

\noindent where the internal angles relate to each other through
Snell's law: $\sin\alpha_0=n_1\sin(\alpha_1-\gamma_a)$ and
$n_1\sin\alpha_1=n_2\sin\alpha_2$ (see Figure \ref{fig5}).

It can be shown for the plane-parallel grating\footnote{After noting
that $\Delta\lambda=2(\sigma/m)\cos\beta\sin\delta$.}, using Equation
\ref{fullR}, that the number of resolution elements on the detector is

\begin{equation}
\label{nresel}
N_R = \frac{\Delta\lambda}{\delta\lambda} =
\frac{2\sin\delta}{r}\frac{f_\mathrm{coll}}{w}
\end{equation}

\noindent where $\Delta\lambda$ is the full wavelength coverage on the
detector.

The following sections discuss two cases, those of a plane-parallel,
tilted fringe grating, and a prism-immersed, untilted fringe grating.

\subsubsection{Tilted Fringe Gratings}

In this section, we will consider plane-parallel gratings
($\gamma_a=\gamma_b=0$), with tilted fringes ($\phi\neq0$). We define
the sign of the tilt such that positive tilts move the plane of the
fringes toward the incident beam and decreases $\beta$; negative tilts
move away and increase $\beta$ (see Figure \ref{fig5}). These gratings
would be most efficient at the Bragg wavelength, not the Littrow
wavelength, and hence would be used in an off-Littrow configuration.
With a non-zero fringe tilt, the Bragg wavelength, $\lambda_B$, is
found to be

\begin{equation}
\label{braggwave}
\lambda_B=\frac{2\sigma n_2}{m}\sin(\alpha_2-\phi)\cos\phi
\end{equation}

\noindent where $n_2$ is the index of refraction of the DCG.  In the
$\phi=0$ case, the Bragg wavelength and the Littrow wavelength
(Equation \ref{littwave}) are the same.  Inserting the above into the
grating equation, one can determine the angle of diffraction of the
Bragg wavelength to be

\begin{equation}
\sin\beta_B = n_2 \sin\left[\arcsin\left(\frac{\sin\alpha}{n_2}\right)
- 2\phi\right]
\end{equation}

\noindent Because this angle is different than that for the Littrow
wavelength, the ghost can be moved off of the detector by introducing
a sufficiently large fringe tilt (positive or negative) such that
$\Delta\beta\equiv\mid\beta_L-\beta_B\mid=\mid\alpha-\beta_B\mid$ is
greater than the relevant fraction of the camera field-of-view.

\begin{figure}[h]
\epsscale{1}
\plotone{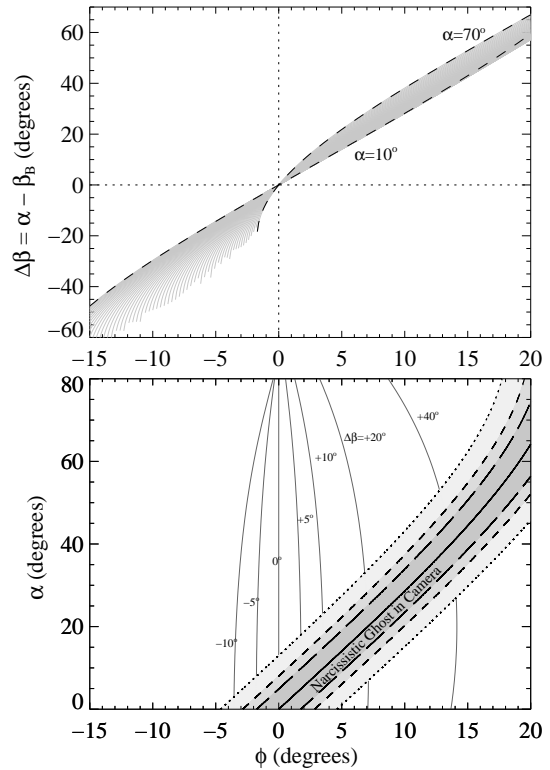}
\caption{ 
Critical angles in tilted-fringe VPH gratings. Top panel: Difference
in $\beta$ between Littrow and Bragg wavelengths for VPH gratings with
tilted fringes as a function of fringe tilt, $\phi$. The relation is
nearly linear and roughly constant with $\alpha$. Two values of
$\alpha$ are overplotted (dashed lines), with intermediate range of
$\alpha$ filled in gray.  When $\Delta \beta \sim \alpha$, the
narcissistic ghost enters the camera.  Bottom panel: Incidence angle
$\alpha$ versus fringe tilt. Shaded regions show where a fringe tilt
would place the narcissistic ghost on the detector for a given
$\alpha$ and $\phi$, assuming the grating is used at the Bragg
angle. The thick solid line is for $\beta_B=0$, and the others for
$\beta_B=0\pm\delta$ for the Bench Spectrograph (long dashed), RSS
(short dashed), and RSS-MOS (dotted). Thin solid lines show constant
$\Delta \beta$ for a range of values. A tilted-fringe VPH grating
should be used over a range of $\alpha$ that avoids the shaded
region. Sufficiently negative fringe tilts avoid the narcissistic
ghost completely.
\label{fig6}}
\end{figure}

The upper panel of Figure \ref{fig6} shows the relationship between
$\Delta\beta$ and $\phi$.  It is linear and roughly independent of
$\alpha$. At angles of $\alpha$ up to 45$\degr$ and $n_2\sim1.4$, the
small angle approximation for the $\sin$ and $\arcsin$ are quite good,
and we can write $\beta_B\simeq\alpha-2n_2\phi$. So, for a given
$\alpha$, the fringes need to be tilted by
$\phi\simeq\Delta\beta/2n_2$ to move by $\Delta\beta$ degrees away
from the position of the Littrow ghost.  The sense of the
approximation is conservative: for larger $\alpha$ this formula gives
an overestimate for the necessary fringe tilt.

For long-slit spectrographs, $\Delta\beta$ should be at least half the
camera FOV, or, as mentioned in Section \ref{off-littrow}, enough to
move the ghost into a detector gap.  For RSS and the Bench
Spectrograph, the camera FOVs are 16$\degr$ and 9$\fdg$8 requiring
$\phi\ge2\fdg8$ and $\phi\ge1\fdg7$, respectively, for $n_2=1.4$, to
move the ghost off of the detector over the full range of conceivable
incident angles. These are very modest tilts.

To ensure that all ghosts are removed for multi-object spectrographs,
the $\Delta\beta$ should be larger to exclude the ghosts from off-axis
objects.  For RSS an additional $\sim5\degr$ needs to be accommodated,
resulting in $\phi\ge4\fdg6$.  However, for configurations with small
grating angles ($\alpha \lesssim 20\degr$), a tilt of the fringes this
large moves $\beta_B$ within a half camera FOV of $\beta_B=0$, the
condition for the appearance of the narcissistic ghost\footnote{The
$\beta=0$ condition occurs when
$\phi=1/2\arcsin\left(\frac{\sin\alpha}{n_2}\right)\approx\alpha/2n_2$.}.
This may be resolved by tilting the fringes in the negative direction,
which moves $\beta_B$ in the opposite sense. Fringe tilts in either
direction have an impact on performance in terms of the merit
functions we described above. Hence, a careful consideration of
Littrow-ghost removal, avoidance of introducing the narcissistic
ghost, and impact on performance merit-functions must be considered
and properly balanced for one's given science goals.

\begin{figure*}[t]
\epsscale{1.08}
\plotone{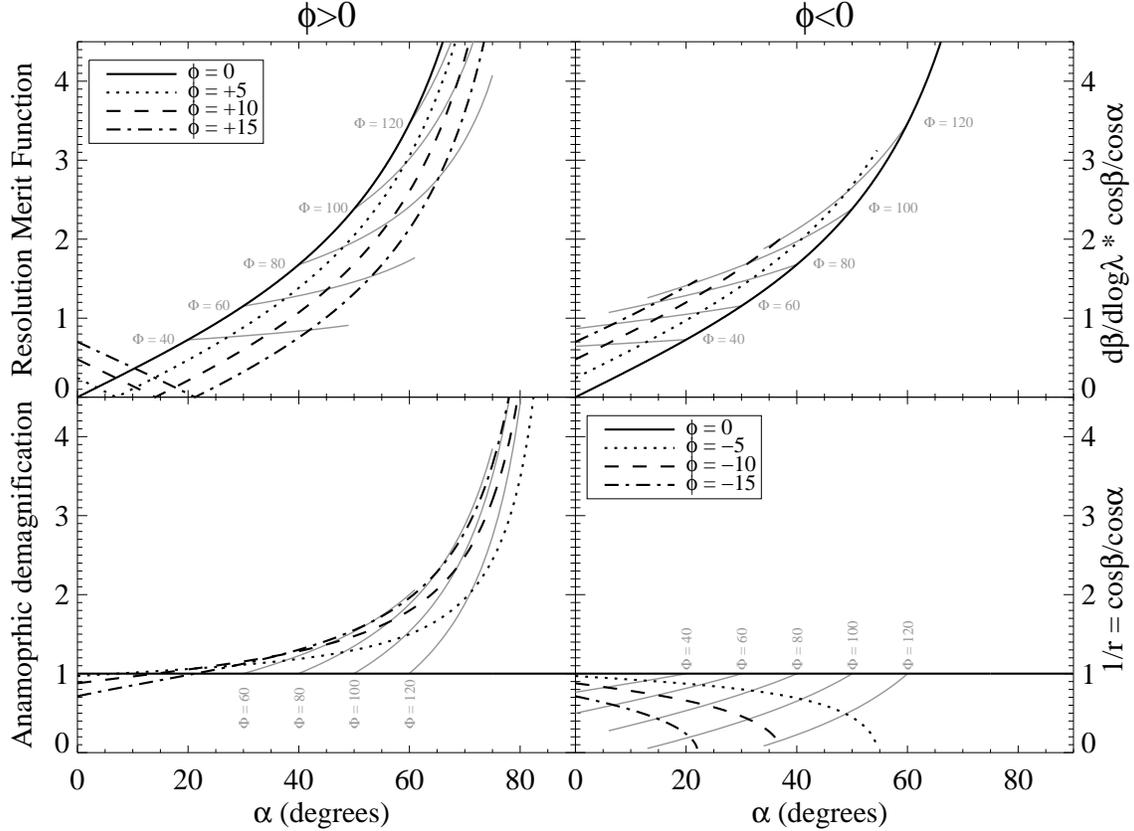}
\caption{Resolution merit function and anamorphic factor evaluated at
the Bragg wavelength for plane-parallel VPH gratings with tilted
fringes and a mean DCG index $n_2=1.4$.  Left panels are for $\phi>0$
and Right panels for $\phi<0$.  Lines of constant camera-collimator
angle, $\Phi=\alpha+\beta$, are shown in gray.  For a fixed $\Phi$,
gratings with positive fringe tilts decrease $\beta$, such that the
accompanying increase in $\alpha$ can produce significant enhancements
in the performance metrics.  However, at small $\alpha$, negative
fringe tilts enhance the resolving power with only modest losses in
spectral coverage due to increased anamorphic magnification, while
preventing the introduction of narcissistic ghosts.
\label{fig7}}
\end{figure*}

Figure \ref{fig7} shows the resolution merit function and the
anamorphic factor (resolution-element merit function) calculated at
the Bragg wavelength versus grating incidence angle for positive and
negative fringe tilts.  Negative fringe tilts give a small amount of
increased resolving power at a given $\alpha$ by significantly
increasing dispersion, which overcomes an increase in the anamorphic
magnification.  However, this means that the detector is less
efficiently used because of the fewer number of resolution elements.
Negative fringe tilts also limit the usable range of $\alpha$ for
which $\beta_B<90\degr$ (transmission) and thus the maximum
achievable resolving power in transmission is lowered.

With positive fringe tilts, the anamorphic demagnification ($1/r$)
increases strongly at large incidence angles (lower left panel of
Figure \ref{fig7}), although there is little gain for
$\phi>15\degr$.  Note that the demagnification becomes $<1$
(i.e., magnification) roughly when $\alpha\sim n_2\phi$.  This is when
the effective diffraction angle ($\alpha_2-\phi$) changes sign with
respect to the tilted fringes (the grating remains in transmission).
The overall resolving power decreases with increased positive fringe
tilt, but the decrease is modest for small tilt angles.  However, the
large increase in anamorphic demagnification increases the number of
resolution elements for a definite gain in information.  The loss in
resolving power can easily be recovered by increasing $\alpha$ and
modulating $\sigma$ in the grating design to tune the wavelength.

In summary, for gratings used at small angles ($\alpha \lesssim20
\degr$) with wide-field spectrographs, modest negative fringe-tilts
should be used to remove Littrow ghosts. Negative tilts will avoid
narcissistic ghosts, and boost the resolution merit-function via
increased dispersion, with some loss in spectral coverage due to
increased anamorphic magnification. The other option, namely to use
positive fringe tilts large enough to avoid narcissistic ghosts, will
also remove Littrow ghosts, but does not do as well in terms of
performance metrics for these low angles.  At larger $\alpha$ it is
beneficial to choose positive tilts in terms of the merit
functions. Because of the narcissistic ghosts, however, the tilt must
be chosen carefully with the range of $\alpha$ in mind (see bottom
panel of Figure \ref{fig7}).

\subsubsection{Prism-Immersed Gratings}

In the previous section we considered a VPH grating sandwiched between
plane-parallel substrates.  By sandwiching the grating between prisms,
the grating incident angle as well as the camera-collimator angle may
be reduced for a given grating and wavelength.  The reduction of these
angles is favorable because air-glass interface losses are smaller at
lower incident angles and there will be physical constraints on the
camera-collimator angle of many spectrographs.  

Figure \ref{fig5} shows the geometry for a general
prism-immersed VPH grating, with prism angles $\gamma_a$ on the
collimator side and $\gamma_b$ on the camera side of the grating.  The
total beam deviation will be $\alpha_0+\beta_0+\gamma_a+\gamma_b$.  In
the case for untilted fringes, $\phi=0$, the merit function reduces to

\begin{equation}
\frac{1}{r}\frac{d\beta}{d\log\lambda} =
\frac{\cos(\alpha_1-\gamma_a)}{\cos\alpha_0}2n_1\tan\alpha_1
\end{equation}

\noindent consistent with Equation A8 of \citet{Baldry04}.  It should
be noted that this holds true regardless of the value of $\gamma_b$,
not just for the symmetric case.

\begin{figure}
\epsscale{1}
\plotone{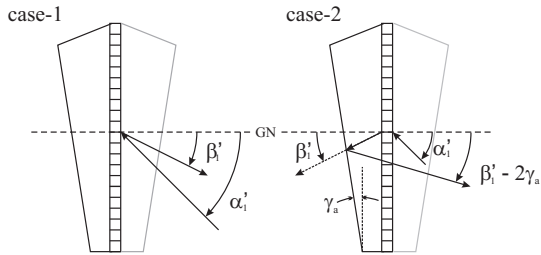}
\caption{Light paths for the case-1 reflective (left) and case-2
transmissive (right) recombinations in a prism-immersed VPH grating.
Only the internal angles are shown.  The transmissive recombination
ghost is redirected by twice the collimator-side prism angle,
$\gamma_a$.  The internal angles are related to the external angles by
$\sin\beta_0=n_1\sin(\beta_1-\gamma_b)$.
\label{fig8}}
\end{figure}

Because of the angle on the prism on the collimator side, the beam
from the transmissive recombination will be redirected upon reflection
so that it does not follow the same path as the Littrow wavelength.
This manifests itself as a change in $\beta_1$ such that $\beta_1^T =
\beta_1-2\gamma_b$ (see Figure \ref{fig8}).  However, the path
of the reflective recombination ghost will remain unchanged.  Thus,
the use of prism-immersed gratings will not completely resolve the
ghost issue, as evidenced by the existence of the ghost in VPH grism
spectrographs.  The only way to remove the reflective recombination
ghost from the camera is to operate the grating itself, i.e., the DCG,
in an off-Littrow configuration, as in the tilted fringe case
discussed in the previous section.

\section{Summary}

VPH gratings are becoming much more common in astronomical
spectrographs.  Despite the clear advantages of using VPH gratings,
the necessity of employing a Littrow configuration for gratings with
untilted fringes produces an optical ghost that may overlap sensitive
sections of the primary spectrum. We have identified two paths for the
Littrow ghost with VPH gratings, involving light reflected off of the
detector surface that is recollimated by the camera and recombined by
the grating before being sent back to the detector.  The efficiencies
of the two paths depend on the specific properties of the grating
and/or anti-reflection coating of the grating substrates, but even
with low efficiency, the great enhancement by full recombination of
the $\Delta m=0$ case produces a ghost that can be as bright or
brighter than the features in the direct spectrum on which it lands.

We have created a theoretical model for the general phenomenon of this
ghost capable of explaining the observations made by the VPH
spectrographs designed and built by the University of
Wisconsin~-~Madison and other institutions as well as those made with
conventional surface-relief gratings.  We have shown that our model is
capable of accurately reproducing the ghost brightness as observed
with reasonable assumptions about the spectrograph and grating
performances.

For existing VPH grating spectrographs there are multiple options to
mitigating the effects of the ghost, including changes to the baseline
spectrograph configuration or dithering between images - either by
changing the spectrograph configuration or by repointing the telescope
to move the source along the slit.  The extent to which any of these
solutions is desirable will depend on the science goals of the
observation as well as the specific spectrograph and telescope
limitations.

Although these solutions may be straightforward for the case of
single-object longslit observations, the removal of the effects of the
ghost for multi-object, fiber-fed and nod-and-shuffle modes may be
complicated.  In MOS modes, telescope nods will be limited because of
the typically short lengths of the individual slits, so unless the
sources are small and dominate the ghost flux, it may be best to
dither the grating/camera angle.  For fiber-coupled systems it will be
difficult to move extended sources (covering multiple fibers, e.g., in
an IFU) properly along the pseudo-slit unless careful attention has
been paid to the telescope-to-spectrograph focal-surface mapping.  The
nod-and-shuffle procedure is typically used for observations of
background-limited sources.  In this case, the ghost will essentially
be similar to a sky line, albeit without line curvature, and as such
is as readily removed as the other sky lines.

For future VPH spectrographs, we discussed solutions for removing the
ghost completely with new grating designs.  We considered both
prism-immersed gratings as well as plane-parallel gratings with tilted
fringes.  We conclude that the latter is superior because of its
ability to remove the ghosts from both the generation paths we have
identified.  We have analyzed the impact of tilted-fringe gratings in
terms of merit functions for spectral resolution and spectral
information, whose product is equivalent to the comprehensive measure
of "spectral power" \citep{Bershady07}.  We find that modest fringe
tilts of $\pm5-15\degr$, sufficient to remove Littrow ghosts, can
significantly improve spectrograph performance. While some care is
needed in the grating design such that the intended range of user
angles ($\alpha$) avoids the introduction of narcissistic ghosts, the
gains are particularly impressive for systems with limited
articulation. For example, a fringe-tilt of only $5\degr$ for RSS
yields up to 80\% more spectral power than the existing, untilted
gratings. This tilt is sufficiently modest that even with the concerns
noted by \citet{Rallison92} about the effects from sag during DCG
processing it is reasonable to expect that there will be little to no
negative impact on performance.  Furthermore, the incident-angles
remain small enough ($\alpha\lesssim60\degr$) for high-performance AR
coatings.  The framework presented here should allow the community to
usher in a new wave of VPH gratings free of ghost artifacts and
boosted in spectral performance by factors of 50 to several 100\%.

\begin{acknowledgements}
We would like to thank Patrick Morrissey for pointing out the
possibility of the transmissive recombination method for the ghost,
and Gary Bernstein for providing his RCWA code and advice on computing
reflection modes. Gratitude is also due to Mike Smith, the chief
mechanical engineer for the RSS; Sam Gabelt, Don Michalski and the
staff of SAL; the University of Wisconsin Physics Department machine
shop; the SAAO staff for their work on the RSS detector subsystem;
Charles Harmer, Di Harmer, Gene McDougall, Pat Knezek, George Jacoby,
Gary Poczulp, and Skip Andre at WIYN/NOAO for technical support; and
Sam Barden for initiating the NOAO VPH program. We also thank Wasatch
Photonics (R. Rallison and D. Cifelli) and CSL (P.-A. Blanche,
P. Lemaire, and S. Habraken) for their efforts making science-grade
gratings for astronomical application. MAB and KBW were supported by
NSF/AST-0307417 and NSF/AST-0607516.

\end{acknowledgements}

\clearpage

\end{document}